\documentclass[aps,prd,preprint,preprintnumbers,unsortedaddress,superscriptaddress,showpacs,nofootinbib]{revtex4-1}

\usepackage{graphicx,color}
\usepackage{relsize}
\usepackage{slashed}
\usepackage{color}
\usepackage{ulem}
\usepackage{tabu}
\usepackage{bm}

\usepackage{amsmath}
\usepackage{amssymb}
\usepackage{amsthm}
\usepackage{mathrsfs}
\usepackage{graphicx}
\usepackage{epstopdf}
\usepackage{fancyhdr}
\usepackage{array}
\usepackage[all]{xy}
\usepackage{eufrak}
\usepackage{euscript}
\usepackage{enumerate}
\usepackage{slashed}
\usepackage{hyperref}
\hypersetup{pdftex,colorlinks=true,linkcolor=blue,citecolor=blue,menucolor=black,urlcolor=blue,filecolor=blue}

\begin{document}

\title{Triangle singularity enhancing isospin violation in $ \boldsymbol{\bar B_s^0 \to J/\psi  \pi^0 f_0(980)}$}

\author{Wei-Hong Liang}
\email{liangwh@gxnu.edu.cn}
\affiliation{Department of Physics, Guangxi Normal University, Guilin 541004, China}
\affiliation{Departamento de F\'{i}sica Te\'{o}rica and IFIC, Centro Mixto Universidad de Valencia - CSIC,
Institutos de Investigaci\'{o}n de Paterna, Aptdo. 22085, 46071 Valencia, Spain.}

\author{S.~Sakai}
\email{shuntaro.sakai@ific.uv.es}
\affiliation{Departamento de F\'{i}sica Te\'{o}rica and IFIC, Centro Mixto Universidad de Valencia - CSIC,
Institutos de Investigaci\'{o}n de Paterna, Aptdo. 22085, 46071 Valencia, Spain.}

\author{Ju-Jun~Xie}
\email{xiejujun@impcas.ac.cn}
\affiliation{Institute of Modern Physics, Chinese Academy of Sciences, Lanzhou 730000, China}
\affiliation{Departamento de F\'{i}sica Te\'{o}rica and IFIC, Centro Mixto Universidad de Valencia - CSIC,
Institutos de Investigaci\'{o}n de Paterna, Aptdo. 22085, 46071 Valencia, Spain.}

\author{E.~Oset}
\email{eulogio.oset@ific.uv.es}
\affiliation{Departamento de F\'{i}sica Te\'{o}rica and IFIC, Centro Mixto Universidad de Valencia - CSIC,
Institutos de Investigaci\'{o}n de Paterna, Aptdo. 22085, 46071 Valencia, Spain.}

\date{\today}

\begin{abstract}
  We perform calculations for the $\bar B_s^0 \to J/\psi \pi^0 f_0(980)$ and $\bar B_s^0 \to J/\psi \pi^0 a_0(980)$ reactions, showing that the first one is isospin-suppressed while the second one is isospin-allowed. The reaction proceeds via a triangle mechanism, with $\bar B_s^0 \to J/\psi K^* \bar K +c.c.$, followed by the decay $K^* \to K\pi$ and a further fusion of $K\bar K$ into the $f_0(980)$ or $a_0(980)$. We show that the mechanism develops a singularity around
  the $\pi^0 f_0(980)$ or $\pi^0 a_0(980)$ invariant mass
  of 1420 MeV where the $\pi^0 f_0$ and $\pi^0 a_0$ decay modes are magnified and also the ratio of $\pi^0 f_0$ to $\pi^0 a_0$ production. Using experimental information for the $\bar B_s^0 \to J/\psi K^* \bar K +c.c.$ decay, we are able to obtain absolute values for the reactions studied which fall into the experimentally accessible range. The reactions proposed and the observables evaluated, when contrasted with actual experiments should be very valuable to obtain information on the nature of the low lying scalar mesons.
\end{abstract}

%\pacs{Valid PACS appear here}
% PACS, the Physics and Astronomy Classification Scheme.
% Valid PACS numbers may be entered using the \verb+\pacs{#1} command.

%\keywords{Baryons, Mesons, Resonances, Molecular states, Chiral unitary approach, Nonperturbative technique.}

\maketitle
%\tableofcontents

%%%%%%%%%%%%%%%%%%%%%%%%%%%%%%%%%%
\section{Introduction}
\label{sec:intro}

Triangle singularities (TS) are capturing the attention of hadron physics (see talk in the latest hadron conference \cite{guosal}). Introduced by Landau in 1959 \cite{Landau:1959fi}, the TS stems from a mechanism that can be represented by a Feynman diagram with a loop with three propagators. An external particle $A$ decays into two particles $1$ and $2$. Particle $2$ decays into particle $3$ and an external particle $B$, and then particles $1$ and $3$ merge into an external particle $C$. The loop contains the particles $1,2,3$ as internal particles. Under certain circumstances which correspond to having the possibility of the process occurring at the classical level, a singularity in the amplitude develops \cite{Coleman:1965xm}. This occurs when all the intermediate particles are placed on-shell and are colinear. The amplitude becomes infinite if the internal particles have zero width. However, the fact that particle $2$ can decay into $3+B$ implies that it has a width, and the infinite amplitude turns into a finite peak, which can be identified experimentally. A reformulation of the problem, in the light of present computing facilities (at the level of a simple PC), offers a more intuitive and practical approach to this issue \cite{Bayar:2016ftu}. The existence of a singularity for a given mechanism is established by means of a single equation, $q_{\rm on} = q_{a-}$ (see Eq. (18) of Ref.~\cite{Bayar:2016ftu}), where $q_{\rm on}$ is the on-shell momentum of particle $1$ in the decay of $A \to 1+2$, and $q_{a-}$ is the smallest momentum for particle $2$, when $2+3$ merge on shell to give the moving particle $C$, with particles $2$ and $B$ having momenta in opposite directions (this situation allows the Coleman Norton theorem \cite{Coleman:1965xm} to be fulfilled). Clear as the problem is, no experimental examples were found for long time, but the situation has reversed recently. Suggestions to find TS in different reactions were done in Ref.~\cite{Liu:2015taa}. In particular, it was suggested that the peak seen by the COMPASS collaboration that was initially associated to a new resonance, the $a_1(1420)$ \cite{Adolph:2015pws}, was a consequence of a triangle singularity that reinforced the $a_1(1260)$ decay into $\pi f_0(980)$. Detailed calculations clearly reaching this conclusion were done in Refs.~\cite{Ketzer:2015tqa,Aceti:2016yeb}. Similarly, arguments have been given in Ref.~\cite{Debastiani:2016xgg} that the $f_1(1420)$ resonance, catalogued as such in the PDG~\cite{PDG2016}, does not correspond to a resonance, but it is a manifestation of the $f_1(1285)$ decay into $K \bar K^*$, with the ``$\pi a_0(980)$ decay mode'' claimed in Ref.~\cite{Barberis} corresponding to a TS enhanced decay mode of the $f_1(1285)$. Another example is given by the $f_2(1810)$ ``resonance'', which as shown in Ref.~\cite{Xie:2016lvs}, comes naturally from a TS involving $K^* \bar K^*$ production, followed by $K^* \to \pi K$ and $\bar K^* K$ fusing into $a_1(1260)$.

Some awakening to the TS was spurred by the suggestion that the $P_c(4450)$ peak seen by the LHCb collaboration \cite{Aaij:2015tga,Aaij:2015fea} might correspond to a TS \cite{Guo:2015umn,Liu:2015fea}, but the follow-up work in Ref.~\cite{Bayar:2016ftu} showed that for the preferred quantum numbers $J^P=3/2^-, \, 5/2^+$ of the experimental analysis this could not be the explanation.

The TS has also helped to explain some peculiar experimental features of different reactions, like the peak around $\sqrt{s}=2110 ~{\rm MeV}$ of the $\gamma p \to K^+ \Lambda(1405)$ reaction \cite{Moriya:2013hwg}, explained in Ref.~\cite{Wang:2016dtb} through a TS, or the $\pi N^*(1535)$ contribution to the $\gamma p \to \pi^0 \eta p$ reaction \cite{Gutz:2014wit}, also explained through such a mechanism in Ref.~\cite{Debastiani:2017dlz}.
A possible $\phi p$ resonance, the hidden-strange analogue of the $P_c$ state, was investigated in the $\Lambda^+_c \to \pi^0 \phi p$ decay by considering a triangle singularity mechanism~\cite{Xie:2017mbe}, where the obtained $\phi p$ invariant mass distribution agrees with the existing Belle data~\cite{Pal:2017ypp}.
Other examples can be found in a more detailed description in Ref.~\cite{abnormal}.

On the other hand, the issue of isospin violation in production of the $f_0(980)$ or $a_0(980)$ resonances, and their mixing, has been a recurrent topic \cite{Achasov:1979xc,Hanhart:2003pg,Wu:2007jh,Hanhart:2007bd,Roca:2012cv}. While trying to establish a ``$f_0-a_0$ mixing parameter''  from different reactions, the concept had to be abandoned because it was shown that the amount of isospin violation was very much reaction dependent. Particularly, it was shown in Refs.~\cite{Wu:2011yx,Aceti:2012dj} that the large isospin violation in the $\eta(1405) \to \pi^0 f_0(980)$ decay \cite{BESIII:2012aa} was due to a TS. Since then, a search for TS enhanced isospin-violating reactions producing the $f_0(980)$ or $a_0(980)$ resonances has been initiated. In Ref.~\cite{Aceti:2015zva} the $f_1(1285)$ decays into the isospin-allowed $\pi^0 a_0(980)$ mode and the isospin-forbidden $\pi^0 f_0(980)$ mode were studied and the latter was confirmed a few months later in a BESIII experiment \cite{Ablikim:2015cob}.
More recently the $D_s^+ \to \pi^+ \pi^0 a_0(980) (f_0(980))$ reaction has been suggested as an example of isospin violation ($D_s^+ \to \pi^+ \pi^0 f_0(980)$) enhanced by a TS \cite{abnormal}. In this reaction, the $D_s^+$ decays into $\pi^+$ and a quark pair $s\bar s$ which hadronizes in two mesons in isospin $I=0$. The TS emerges from the decay mode $D_s^+ \to \pi^+ (K^+ K^{*-} + K^0 \bar K^{*0})$ followed by $\bar K^* \to \pi^0 \bar K$ and $K\bar K$ merging into the $a_0(980)$ (the isospin-allowed mode). The mechanism produces a TS at around $1420 ~{\rm MeV}$ of the invariant mass of $\pi^0 a_0(980)$, $M_{\rm inv}(\pi^0 a_0(980))$ . The isospin-forbidden $D_s^+ \to \pi^+ \pi^0 f_0(980)$ mode emerges from the lack of cancellation between the $K^0 \bar K^0$ and $K^+ K^-$ intermediate states in the loops, and it is shown that the mode is enhanced with respect to the isospin-allowed mode around the TS peak.

Following this line of research, in this work we present a different reaction, $\bar B_s^0 \to J/\psi \pi^0 f_0(980) (a_0(980))$, in which the $f_0(980)$ production mode is also isospin-forbidden. The reaction has  different dynamics than the $D_s^+ \to \pi^+ \pi^0 f_0(980) (a_0(980))$ but shares some features concerning the TS. We also observe an enhancement of ${\rm d} \Gamma / {\rm d} M_{\rm inv}(\pi^0 f_0)$ and ${\rm d} \Gamma / {\rm d} M_{\rm inv}(\pi^0 a_0)$ around $M_{\rm inv}= 1420~{\rm MeV}$, and the ratio of these two distributions also peaks around this value of the invariant mass. These features are tied to the picture of the $f_0(980)$ and $a_0(980)$ as dynamically generated states from the interaction of pseudoscalar mesons, and their experimental confirmation will be relevant to gain further insight into the nature of the low lying scalar mesons.

%%%%%%%%%%%%%%%%%%%%%%%%%%%%%%%%%%
\section{Formalism}
\label{sec:form}
\subsection{The $\boldsymbol{\bar B_s^0 \to J/\psi K^{*0} \bar K^0}$ decay}
\label{subsec:K0-amplitude}

\begin{figure}[b!]
\begin{center}
\includegraphics[scale=0.6]{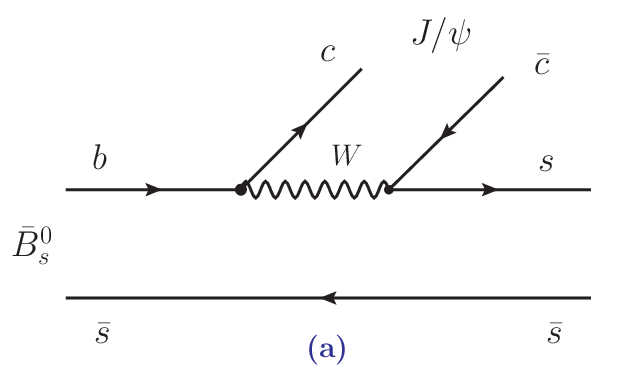}
\hspace{0.5cm}
\includegraphics[scale=0.6]{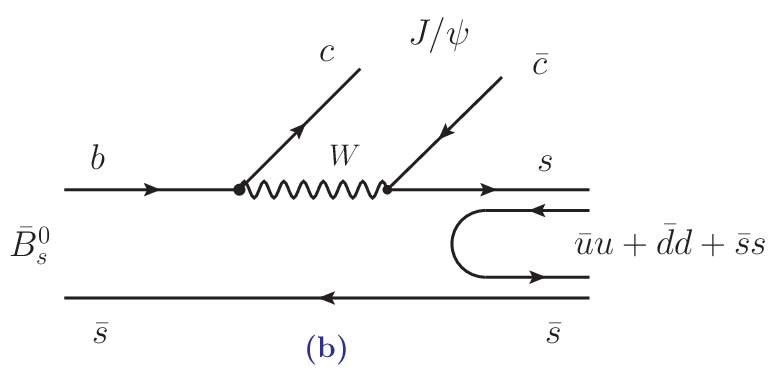}
\end{center}
\vspace{-0.7cm}
\caption{Diagrammatic representation of $\bar B^0_s \to J/\psi (c\bar c) s\bar s$ at the quark level.}
\label{Fig:1}
\end{figure}

We describe the $\bar B_s^0 \to J/\psi \pi^0 f_0(980) (a_0(980))$ reaction. In a first step we show in Fig. \ref{Fig:1}(a) the basic decay of $\bar B_s^0$ into $J/\psi (c\bar c)$ and a pair of quarks $s\bar s$. This mechanism proceeds via internal emission \cite{Chau:1987tk,LiangPLB2014}, and leaving apart the $bcW$ vertex, needed for the decay, the second vertex, $Wcs$, is Cabibbo favored. The next step consists of the hadronization of $s\bar s$ to give a pair of mesons, which is shown in Fig. \ref{Fig:1}(b).
Following the step of Refs. \cite{LiangPLB2014,Oset:2016lyh}, we can write
\begin{equation}\label{Eq:qqbar}
  s(\bar u u+\bar d d +\bar ss) \bar s =\sum_{i=1}^3 \mathcal{M}_{3i} \mathcal{M}_{i3},
\end{equation}
where $i$ runs over the quarks $u,d,s$, and $\mathcal{M}$ is the $q \bar q$ matrix in SU(3). We can write the $\mathcal{M}$ matrix in terms of pseudoscalar mesons, $\Phi$, or vector mesons, $V$, as
\begin{equation}
\label{Phimatrix}
\Phi =
\left(
\begin{array}{ccc}
 \frac{1}{\sqrt{2}} \pi^0 +\frac{1}{\sqrt{3}} \eta  + \frac{1}{\sqrt{6}} \eta^{\prime} & \pi^+ & K^+ \\
\pi^- & ~- \frac{1}{\sqrt{2}} \pi^0 + \frac{1}{\sqrt{3}} \eta  + \frac{1}{\sqrt{6}} \eta^{\prime} ~& K^0 \\
K^- & \bar{K}^0 & -\frac{1}{\sqrt{3}} \eta + \sqrt{\frac{2}{3}} \eta^{\prime}  \\
\end{array}
\right)\, ,
\end{equation}

\vspace{0.3cm}
\begin{equation}
\label{Vmatrix}
V =
\left(
\begin{array}{ccc}
\frac{1}{\sqrt{2}} \rho^0 + \frac{1}{\sqrt{2}} \omega & \rho^+ & K^{* +}  \\
\rho^- & ~-\frac{1}{\sqrt{2}} \rho^0 + \frac{1}{\sqrt{2}} \omega ~& K^{* 0} \\
K^{* -} & \bar{K}^{* 0} & \phi  \\
\end{array}
\right)\, .
\end{equation}
For reasons that will become clear later, we choose for one $\mathcal{M}$ the matrix $\Phi$ and for the other the matrix $V$ and we get the possible combinations for $s(\bar u u+\bar d d +\bar ss) \bar s$,
\begin{equation}
  K^- K^{*+} + \bar K^0 K^{*0} + \left( - \frac{1}{\sqrt{3}}\eta + \sqrt{\frac{2}{3}} \eta' \right) \phi,
\end{equation}
or
\begin{equation}\label{eq:combin1}
 K^{*-} K^+  + \bar K^{*0} K^0  + \phi \left( - \frac{1}{\sqrt{3}}\eta + \sqrt{\frac{2}{3}} \eta' \right).
\end{equation}
In the triangle diagram that we shall discuss briefly, the $\bar K K^*$ or $\bar K^* K$ will convert into $\pi^0 f_0$ or $\pi^0 a_0$, which have $C$-parity positive. This means that in order to get this final state we must take the $C$-parity positive combination of $\bar K K^*$ and $\bar K^* K$, which under the implicit prescription $C K^*=-\bar K^*$ that we use is given by
\begin{equation}\label{eq:combin2}
  K^- K^{*+} + \bar K^0 K^{*0}-K^{*-} K^+  - \bar K^{*0} K^0,
\end{equation}
and the process that we are interested in is
\begin{equation}\label{eq:process}
  \bar B^0_s \to J/\psi \, (K^- K^{*+} + \bar K^0 K^{*0}-K^{*-} K^+  - \bar K^{*0} K^0).
\end{equation}
The strength of this process is obtained by using the experimental branching ratio for
\begin{equation*}
  B^0_s \to J/\psi K^0 K^- \pi^+ + c.c.,
\end{equation*}
which has a branching fraction \cite{PDG2016,lhcbex}
\begin{equation}\label{eq:BR}
  {\mathrm{Br}}(B^0_s \to J/\psi K^0 K^- \pi^+ + c.c.) = (9.3 \pm 1.3)\times 10^{-4}.
\end{equation}
In the experiment of Ref. \cite{lhcbex}, the $K^0 \pi^+$
or $K^- \pi^+$ are both producing the $K^{*+}, \bar K^{*0}$, from where
one concludes that the rate for $B^0_s \to J/\psi K^0 \bar K^{*0}$ is
one fourth of the rate of Eq.~\eqref{eq:BR}, since the complex conjugate
part of Eq.~\eqref{eq:BR} equals the rate of $B^0_s \to J/\psi K^0 K^-
\pi^+$. Since we are interested in the strength of the amplitude for the
process of Eq.~\eqref{eq:process} with $K\bar K^*, \bar K K^*$ having
$C$-parity positive, we assume that both $C$-parity positive and
negative would give the same contribution ({we shall come back to this point}) and then conclude that
\begin{equation}\label{eq:BR-2}
  {\mathrm{Br}}\left( B^0_s \to J/\psi K^0 \bar K^{*0} (K^- \pi^+)\right)_{C=+} = \frac{1}{8}(9.3 \pm 1.3)\times 10^{-4}.
\end{equation}
But, since $K^{*0} \to K^+ \pi^-, K^0 \pi^0$ with strengths $\frac{2}{3}$, $\frac{1}{3}$ respectively, we have
\begin{equation}\label{eq:BR-3}
  {\mathrm{Br}}\left( B^0_s \to J/\psi K^0 \bar K^{*0}\right) =\frac{3}{2} \frac{1}{8}(9.3 \pm 1.3)\times 10^{-4}.
\end{equation}
We also take the structure for the amplitude of this decay, suited to
the production of two vectors, as in Refs. \cite{Pavao:2017kcr,Sakai:2017hpg}
\begin{equation}\label{eq:tmatrix-1}
  t_{\bar B^0_s \to J/\psi \bar K^0 K^{*0}} =\mathcal{C} \; \epsilon_\mu (J/\psi) \; \epsilon^\mu (K^*).
\end{equation}
As usual, we take the lowest possible angular momentum, but we
shall check the consistency later.
The coefficient $\mathcal{C}$ is obtained by comparing the strength of
Eq.~\eqref{eq:BR-3} with the integral over the invariant
{masses of $J/\psi K^{*0}$ and $K^{*0}\bar{K}^0$. We have \cite{PDG2016}}
\begin{equation}\label{eq:dGamma}
  \frac{{\rm d}^2 \Gamma_{\bar B^0_s \to J/\psi K^{*0} \bar K^0 }
  }{{\rm d} M_{\rm inv}(J/\psi K^{*0}){\rm d} M_{\rm
  inv}(K^{*0}\bar{K}^0)}
  =\frac{M_{\rm inv}(J/\psi K^{*0})M_{\rm inv}(K^{*0}\bar{K}^0)}{(2\pi)^3\,8M_{\bar B_s^0}^3}\, {\overline{ \sum}} \sum \left| t_{\bar B^0_s \to J/\psi  K^{*0} \bar K^0} \right|^2,
\end{equation}

The sum over polarizations of $\left| t_{\bar B^0_s \to J/\psi K^{*0} \bar K^0} \right|^2$ is given by
\begin{equation}\label{eq:t2}
  {\overline{ \sum}} \sum \left| t_{\bar B^0_s \to J/\psi K^{*0} \bar K^0} \right|^2 = \mathcal{C}^2 \left[  2+ \frac{\left( M^2_{\rm inv}(J/\psi K^{*0}) - m^2_{J/\psi} -m^2_{K^{*0}}  \right)^2}{4\, m^2_{J/\psi}\, m^2_{K^{*0}}} \right].
\end{equation}
Thus,
\begin{equation}\label{eq:Cfactor}
  \frac{\mathcal{C}^2}{\Gamma_{\bar B^0_s}}=\frac{{\rm Br}(\bar B^0_s
  \to J/\psi K^{*0} \bar K^0)}{{\displaystyle \int} {\rm d} M_{\rm
  inv}(J/\psi K^{*0})\;{\displaystyle \int} {\rm d} M_{\rm
  inv}(K^{*0}\bar{K}^0)\;
   \frac{1}{\mathcal{C}^2}\frac{{\rm d}^2 \Gamma_{\bar B^0_s \to J/\psi K^{*0} \bar K^0 }
  }{{\rm d} M_{\rm inv}(J/\psi K^{*0}){\rm d} M_{\rm
  inv}(K^{*0}\bar{K}^0)}}.
\end{equation}
{If we want to obtain $\frac{{\rm d} \Gamma_{\bar B^0_s \to J/\psi K^{*0} \bar K^0 }
  }{{\rm d} M_{\rm inv}(J/\psi K^{*0})}$ we integrate the double
  differential width over ${\rm d} M_{\rm inv}(K^{*0}\bar{K}^0)$ and
  conversely, if we wish to get $\frac{{\rm d} \Gamma_{\bar B^0_s \to J/\psi K^{*0} \bar K^0 }
  }{{\rm d} M_{\rm inv}(K^{*0}\bar{K}^0)}$ we integrate the double
  differential width with respect to ${\rm d} M_{\rm inv}(J/\psi
  K^{*0})$.
  The limits of the integration are given by the PDG~\cite{PDG2016}.

  Experimentally we have these two mass distributions in Fig.~10 of
  Ref.~\cite{lhcbex}, and one finds a peak around 1500~MeV in the
  $K^*\bar{K}$ mass distribution, which cannot be obtained from the
  structure of Eq.~\eqref{eq:tmatrix-1}.
  The structure indicates that there is a term like the one in
  Eq.~\eqref{eq:tmatrix-1} and another one that would come from the
  interaction of $K^*\bar{K}$ to give a resonance around 1500~MeV.
  Consistent with the implicit $s$-wave for $K^*\bar{K}$ given by the
  structure of Eq.~\eqref{eq:tmatrix-1}, we take the $f_1(1510)$
  resonance and a structure of the type
  \begin{equation}\label{eq:tmatrix-1p}
   t'_{\bar B^0_s \to J/\psi K^{*0} \bar K^0} =
    \mathcal{C}\;\epsilon_\mu(J/\psi)\;\epsilon^\mu(K^{*})\;F(M_{\rm
    inv}(K^{*0}\bar{K}^0)),
  \end{equation}
  where
  \begin{equation}\label{eq:f}
   F(M_{\rm inv}(K^{*0}\bar{K}^0))=1+a\frac{M_{f_1}\Gamma_{f_1}}{M_{\rm inv}^2(K^{*0}\bar{K}^0)-M_{f_1}^2+iM_{f_1}\Gamma_{f_1}}
  \end{equation}
  and fit the parameter ``$a$'' to get the shape of the experimental
  mass distribution of Ref.~\cite{lhcbex}.
  Then, we have no freedom for $\frac{{\rm d} \Gamma_{\bar B^0_s \to J/\psi K^{*0} \bar K^0 }
  }{{\rm d} M_{\rm inv}(J/\psi K^{*0})}$.

  Taking into account Eqs.~\eqref{eq:tmatrix-1p} and \eqref{eq:f},
  Eq.~\eqref{eq:t2} is replaced by the same one multiplied by $|F(M_{\rm
  inv}(K^{*0}\bar{K}^0))|^2$.

  In Fig.~\ref{fig:mass-dist-primary}, we show both $K^{*0} \bar K^0$ and $J/\psi K^{*0}$ mass distributions compared with experiment.
  \begin{figure}[t]
   \includegraphics[width=8cm]{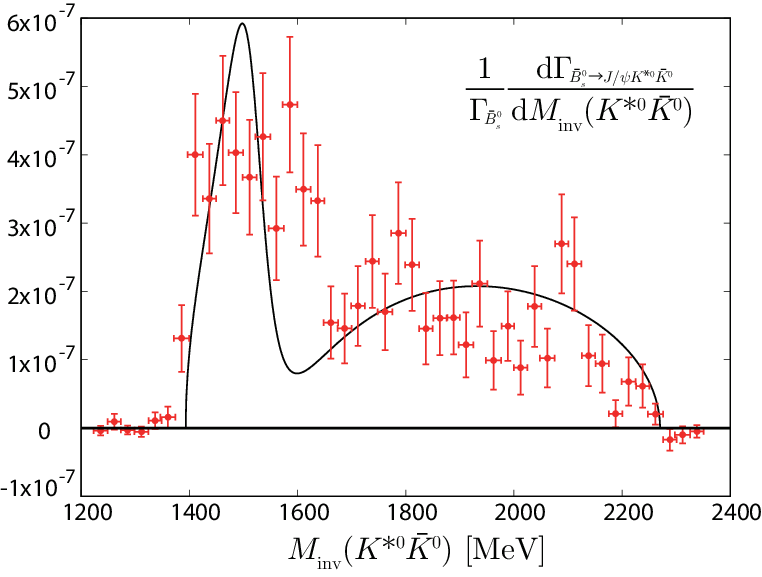}
   \includegraphics[width=8cm]{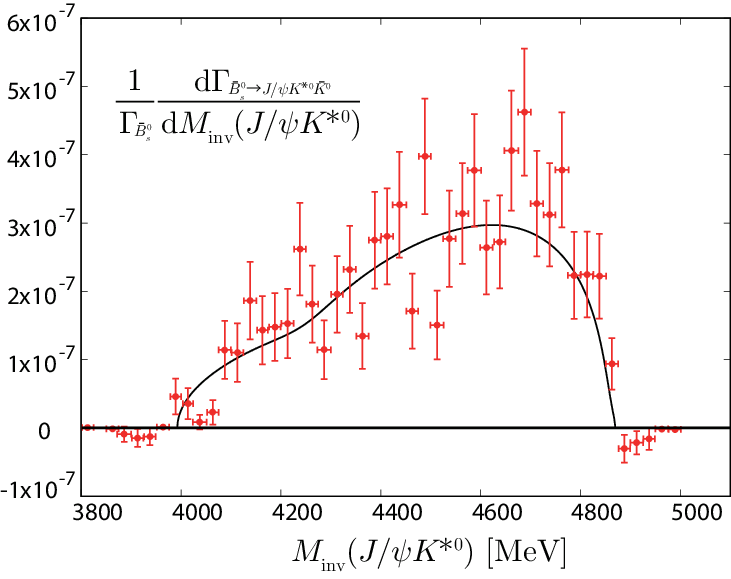}
   \caption{Mass distributions
   ${\rm d}\Gamma_{\bar{B}^0_s\rightarrow J/\psi K^{*0}\bar{K}^0}/{\rm
   d}M_{\rm inv}(K^{*0}\bar{K}^0)$
   and
   ${\rm d}\Gamma_{\bar{B}^0_s\rightarrow J/\psi K^{*0}\bar{K}^0}/{\rm
   d}M_{\rm inv}(J/\psi K^{*0})$
   as functions of $M_{\rm inv}(K^{*0}\bar{K}^0)$ and $M_{\rm
   inv}(J/\psi K^{*0})$, respectively.
   The data are taken from Ref.~\cite{lhcbex} and scaled to agree with
   the calculated mass distribution
   $[{\rm d}\Gamma_{\bar{B}^0_s\rightarrow J/\psi K^{*0}\bar{K}^0}/{\rm
   d}M_{\rm inv}(K^{*0}\bar{K}^0)]/\Gamma_{\bar{B}_s^0}$.}
   \label{fig:mass-dist-primary}
  \end{figure}
  We take $M_{f_1}=1518$~MeV and $\Gamma_{f_1}=98$~MeV compatible with the data
  of the PDG~\cite{PDG2016} and the parameter $a=-1.2$ to agree with the
  data in Ref.~\cite{lhcbex}.
  We see that we account for the bulk of the $K^{*0}\bar{K}^0$ data, and
  the mass distribution of $J/\psi K^{*0}$, which is not fitted, agrees
  well with the data.
  It is clear that the $K^{*0}\bar{K}^0$ mass distribution in
  Fig.~\ref{fig:mass-dist-primary} also has some resonance-like
  structures around 1750~MeV and 2100~MeV, but their strength is much
  smaller than at the peak of 1518~MeV and there is also some extra
  strength around 1600~MeV.
  We neglect these higher resonance contributions, but it is clear that
  we account for most of the strength of the distribution.

  Since the structure proposed provides a reasonable description of the
  data, we can see that the reaction $0^-\rightarrow 1^-1^-0^-$
  ($s$-wave) respects parity.
  Inasmuch as $CP$ is a very good symmetry in weak reactions, if
  parity is conserved, so is $C$ parity.
  Since $\bar{B}_s^0$ is an equal mixture of $CP$ positive and negative,
  we must also expect an equal mixture of $CP$ positive and negative for
  $K\bar{K}^*$ and $\bar{K}K^*$, and with $P$ also conserved, an equal
  mixture of $C$ parity states.
  }

\subsection{Triangle diagram mechanism for $\boldsymbol{\bar B_s^0 \to J/\psi \pi^0 f_0 (a_0)}$ }
\label{subsec:TriangleDiag}

\begin{figure}[b]
\begin{center}
\includegraphics[scale=0.6]{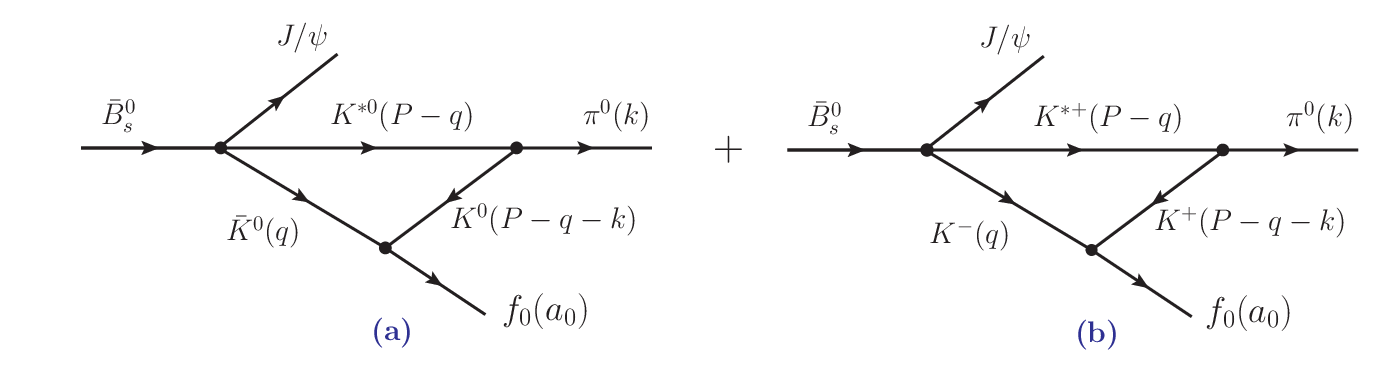}
\hspace{0.5cm}
\includegraphics[scale=0.6]{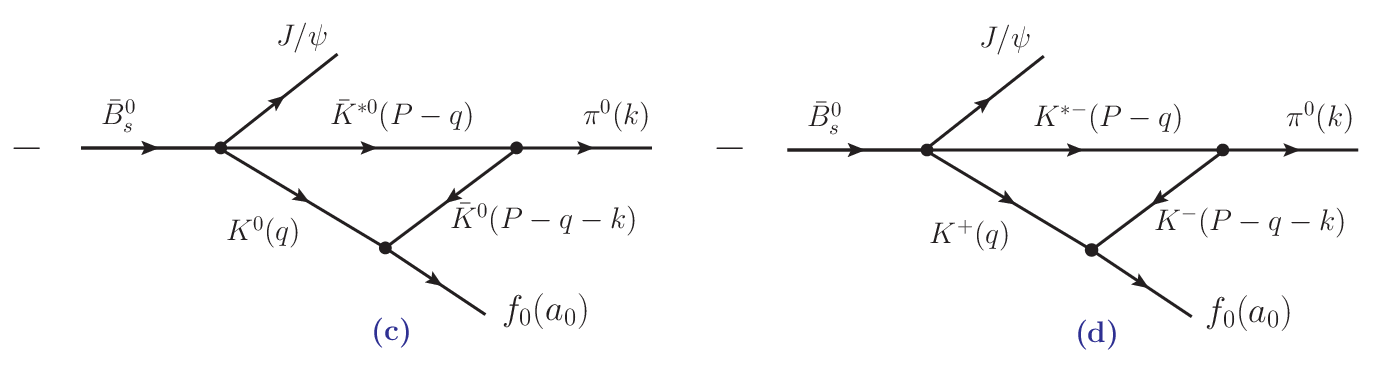}
\end{center}
\vspace{-0.7cm}
\caption{Triangle diagrams for the $\bar B^0_s \to J/\psi \pi^0 f_0 (a_0)$ decay. The parentheses give the momenta of the particles with $P= p_{\bar B^0_s} - p_{J/\psi}$.}
\label{Fig:2}
\end{figure}

In the former subsection we studied on the $\bar B^0_s \to J/\psi K^{*0} \bar K^0$ decay in order to estimate the strength of the transition of Eq.~\eqref{eq:process}. Next we show how the $J/\psi \pi^0 f_0 (a_0)$ is produced using this input. We look now into the related, and unavoidable, mechanism depicted in Fig.~\ref{Fig:2}.
In this mechanism, the $\bar B_s^0$ decays into the $J/\psi \bar K K^*$ (or $\bar K^* K$), the $K^*$ (or $\bar K^*$) decays into $\pi K$ (or $\pi \bar K$), and then the $K$ and $\bar K$ merge to give the $a_0(980)$ or $f_0(980)$ in the final state.

The evaluation of the diagrams requires the use of the $K^* \to K\pi$ amplitude, which comes from the vector(V)-pseudoscalar(P)-pseudoscalar(P) Lagrangian
\begin{equation}\label{eq:Lagrangian}
  \mathcal{L}_{\rm VPP}=-ig \langle \, [\Phi\, , \partial_{\mu}\Phi ] \, V^\mu \, \rangle,
\end{equation}
with $\langle ~\rangle$ the trace in SU(3), $g= \frac{M_V}{2f_\pi} $, $m_V \sim 800 \,{\rm MeV}$ the vector mass, $f_\pi=93 \,{\rm MeV}$ the decay constant of pion, and $\Phi$ and $V$ given by Eqs.~\eqref{Phimatrix}, \eqref{Vmatrix}.
 The $K^{*0} \to \pi^0 K^0$ and $\bar K^{*0} \to \pi^0 \bar K^0$ amplitudes, stemming from Eq.~\eqref{eq:Lagrangian}, have opposite signs, and the same happens with $K^{*+} \to \pi^0 K^+$ and $K^{*-} \to \pi^0 K^-$. Hence, diagrams Fig.~\ref{Fig:2}(a) and~\ref{Fig:2}(c) with the minus sign give the same contribution and so do Fig.~\ref{Fig:2}(b) and~\ref{Fig:2}(d) with the minus sign. Should we have the $C$-parity negative $K^* \bar K$ combination the $(-)$ sign would be replaced by a $(+)$ sign and the diagrams would cancel, as it should be since $\pi^0 f_0$, $\pi^0 a_0$ are $C$-parity positive.

For the amplitude of the diagram of Fig.~\ref{Fig:2}(a) for $\pi^0 a_0$ production, we obtain
\begin{eqnarray}\label{eq:it}
  -it &=& -i \mathcal{C}\; \epsilon_\mu (J/\psi) \, F(M_{\rm inv}(\pi^0a_0)) \int \frac{{\rm d}^4 q}{(2\pi)^4}\; \epsilon^\mu (K^{*0})\; \frac{i}{q^2-m^2_{\bar K^0}+ i\varepsilon} \; \frac{i}{(P-q)^2-m^2_{K^{*0}}+ i\varepsilon}\nonumber \\
   && \cdot  \frac{i}{(P-q-k)^2-m^2_{K^{0}}+ i\varepsilon} \; (-i)  \frac{g}{\sqrt{2}}\; (k-P+q+k)_\nu \; \epsilon^\nu (K^{*0})\; (-i)\, g_{a_0,\, K^0 \bar K^0},
\end{eqnarray}
where $g_{a_0, \,K^0 \bar K^0}$ is the coupling of the $a_0$ resonance
to $K^0 \bar K^0$,
{and $P^0=M_{\rm inv}(\pi^0a_0)$ in the $\pi^0a_0$ rest frame.}

By taking the $a_0(980)$ mass slightly above the $K\bar K$ threshold to apply Eq.~(18) of Ref.~\cite{Bayar:2016ftu}, we find that there is a triangle singularity for this diagram at $M_{\rm inv}(\pi^0 a_0) \sim 1424~{\rm MeV}$. The singularity turns into a finite peak around that mass where most of the strength of the mechanism is concentrated. We take advantage of this fact because then, recalling that the TS places the internal particles on-shell, the on-shell $K^{*0}$ momentum in the loop in the frame of $\pi^0 a_0$ at rest is $163~{\rm MeV}/c$. This allows us to ignore the $\epsilon^0$ component of $K^{*0}$, which only introduces corrections of order $(p_{K^*}/m_{K^*})^2$ with a coefficient that renders this correction smaller than 1\% (see appendix of Ref.~\cite{Sakai:2017hpg}). Then $t$ of Eq.~\eqref{eq:it} becomes
\begin{eqnarray}\label{eq:t-2}
  t &=& \mathcal{C}\; \epsilon_j (J/\psi)\;F(M_{\rm inv}(\pi^0a_0)) \;\, i \int \frac{{\rm d}^4 q}{(2\pi)^4}\; \frac{1}{q^2-m^2_{\bar K^0}+ i\varepsilon} \; \frac{1}{(P-q)^2-m^2_{K^{*0}}+ i\varepsilon}\nonumber \\
   && \cdot  \frac{1}{(P-q-k)^2-m^2_{K^{0}}+ i\varepsilon} \; (2k+q)_j \; \frac{g}{\sqrt{2}}\;  g_{a_0,\, K^0 \bar K^0}.
\end{eqnarray}

Next, as done in Refs.~\cite{Bayar:2016ftu,Aceti:2015zva}, we perform the $q^0$ integration analytically, leaving a ${\rm d}^3 q$ integral to be performed numerically. In addition, since $\vec k$ is the only momentum not integrated in Eq.~\eqref{eq:t-2} (we evaluate $t$ in the rest frame of $\pi^0 a_0$ where $\vec P =0$ ), we can replace $\int {\rm d}^3 q \, \vec q \cdots$ by $\vec k \int {\rm d}^3 q \;\frac{\vec q \cdot \vec k}{{\vec k}^2} \cdots$ and then $t$ of Eq.~\eqref{eq:t-2} can be rewritten as
\begin{equation}\label{eq:t-3}
  t= \mathcal{C}\, \frac{g}{\sqrt{2}}\; g_{a_0,\, K^0 \bar K^0} \; \vec
   \epsilon\,(J/\psi) \cdot \vec k\;F(M_{\rm inv}(\pi^0a_0)) \; t_T,
\end{equation}
with
\begin{eqnarray}\label{eq:tT}
 t_T &=& \int \frac{{\rm d}^3 q}{(2\pi)^3} \; \frac{1}{8\, \omega_{K^0}\, \omega_{K^{*0}} \, \omega_{\bar K^0}} \;
  \frac{1}{k^0-\omega_{K^0}- \omega_{K^{*0}} +i\frac{\Gamma_{K^{*0}}}{2}} \;
  \frac{1}{M_{\rm inv}(\pi^0 a_0) + \omega_{\bar K^0} +\omega_{K^0} -k^0}\; \nonumber \\
   && \times \, \frac{1}{M_{\rm inv}(\pi^0 a_0) - \omega_{\bar K^0}- \omega_{K^0} -k^0 +i\varepsilon} \nonumber \\
  && \times \, \frac{2 M_{\rm inv}(\pi^0 a_0) \, \omega_{\bar K^0}+2k^0 \omega_{K^0} -2(\omega_{\bar K^0}+\omega_{K^0})(\omega_{\bar K^0}+\omega_{K^{*0}}+\omega_{K^0})}{M_{\rm inv}(\pi^0 a_0)-\omega_{K^{*0}}-\omega_{\bar K^0} +i\frac{\Gamma_{K^{*0}}}{2}}\;
  \left(  2+ \frac{\vec q \cdot \vec k}{\left. {\vec k} \right. ^2} \right),~~~
\end{eqnarray}
where
\begin{equation}
 \omega_{\bar K^0}=\sqrt{\vec q\,^2 + m^2_{\bar K^0}},
\end{equation}
\begin{equation}
 \omega_{K^0}=\sqrt{(\vec q + \vec k)^2 + m^2_{K^0}},
\end{equation}
\begin{equation}
 \omega_{K^{*0}}=\sqrt{\vec q\,^2 + m^2_{K^{*0}}},
\end{equation}
\begin{equation}
 k^0=\frac{M^2_{\rm inv}(\pi^0 a_0)+m^2_{\pi^0}-m^2_{a_0}}{2M_{\rm inv}(\pi^0 a_0)},
\end{equation}
\begin{equation}\label{eq:k}
 | \vec k |=\frac{\lambda^{1/2} \left( M^2_{\rm inv}(\pi^0 a_0),\, m^2_{\pi^0}, \,m^2_{a_0}\right)}{2M_{\rm inv}(\pi^0 a_0)}.
\end{equation}

\subsection{Invariant mass distribution}
\label{subsec:invMass}

The invariant mass distribution for $\pi^0 a_0$ is given by
\begin{equation}\label{eq:dGamma}
  \frac{{\rm d} \Gamma}{{\rm d} M_{\rm inv}(\pi^0 a_0)} = \frac{1}{(2\pi)^3} \; \frac{p_{J/\psi} \; \tilde{p}_{\pi^0}}{4\, M^2_{\bar B_s^0}} \; \sum_{\rm pol} |t|^2,
\end{equation}
with
\begin{equation}\label{eq:Sumt2}
 \sum_{\rm pol} |t|^2= \mathcal{C}^2 \; \frac{g^2}{2} \; g^2_{a_0, \,K^0 \bar K^0} \; \left| t_T \right|^2 \, |\vec k|^2 \; \left| F(M_{\rm inv}(\pi^0a_0))  \right|^2,
\end{equation}
and
\begin{equation}\label{eq:pJpsi}
p_{J/\psi} = \frac{\lambda^{1/2} \left( M^2_{\bar B^0_s},\, m^2_{J/\psi}, \, M^2_{\rm inv}(\pi^0 a_0) \right)}{2M_{\bar B^0_s}},
\end{equation}
\begin{equation}
\tilde{p}_{\pi^0} \equiv | \vec k | =\frac{\lambda^{1/2} \left( M^2_{\rm inv}(\pi^0 a_0),\, m^2_{\pi^0}, \, m^2_{a_0} \right)}{2M_{\rm inv}(\pi^0 a_0)}.
\end{equation}

Next we consider that the $a_0$ will be seen in the $\pi^0 \eta$ mass distribution for the decay of the $a_0$ and look at the double differential mass distribution in $M_{\rm inv}(\pi^0 a_0)$ and $M_{\rm inv}(\pi^0 \eta)$. This is done in detail in Ref.~\cite{Pavao:2017kcr} and we write the final result given by
\begin{eqnarray}
   && \frac{1}{\Gamma_{B_s^0}}\; \frac{{\rm d}^2 \Gamma}{{\rm d}M_{\rm inv}(\pi^0 a_0)\; {\rm d}M_{\rm inv}(\pi^0 \eta)} \nonumber \\
  &=& \frac{1}{(2\pi)^5}\; \frac{1}{4 M^2_{B_s^0}} \; p_{J/\psi}\; |\vec k |^3\; \tilde{q}_\eta
  \; \frac{\mathcal{C}^2}{\Gamma_{B_s^0}}\; \frac{1}{2}\,g^2 \left| t_T \right|^2 \; \left| t_{K^0 \bar K^0,\, \pi^0 \eta} \right|^2 \; \left| F(M_{\rm inv}(\pi^0a_0))  \right|^2,
  \label{eq:ddGamma-1}
\end{eqnarray}
where now $\frac{\mathcal{C}^2}{\Gamma_{B_s^0}}$ is taken from Eq.~\eqref{eq:Cfactor}, and $t_{K^0 \bar K^0,\, \pi^0 \eta}$ is the scattering amplitude for $K^0 \bar K^0 \to \pi^0 \eta$ which is calculated using the chiral unitary approach \cite{Oller:1997ti}, but keeping the masses of the $K^0, K^+$ different, which introduces some isospin breaking in the ${\rm P P}\to {\rm P P}$ scattering amplitudes. In Eq.~\eqref{eq:ddGamma-1} the momenta are given by Eqs.~\eqref{eq:k}, \eqref{eq:pJpsi}, replacing $m^2_{a_0}$ with $M_{\rm inv}^2 (\pi^0 \eta)$, and
\begin{equation}
\tilde{q}_{\eta} =\frac{\lambda^{1/2} \left( M^2_{\rm inv}(\pi^0 \eta),\, m^2_{\pi^0}, \, m^2_{\eta} \right)}{2M_{\rm inv}(\pi^0 \eta)}.
\end{equation}

So far we have only considered the contribution of the diagram of Fig.~\ref{Fig:2}(a). We must consider explicitly the contribution of diagram Fig.~\ref{Fig:2}(b) and multiply by two to account for Fig.~\ref{Fig:2}(c) and \ref{Fig:2}(d). This is done replacing $t_T \; t_{K^0 \bar K^0,\, \pi^0 \eta}$ by
\begin{eqnarray}\label{eq:replace}
   t_T \; t_{K^0 \bar K^0,\, \pi^0 \eta} & \longrightarrow & 2 \, \Big\{  t_T \left( \bar K^0,\, K^0,\, K^{*0},\, m_{a_0} \to M_{\rm inv}(\pi^0 \eta) \right) \cdot t_{K^0 \bar K^0,\, \pi^0 \eta}(M_{\rm inv}(\pi^0 \eta))
     \nonumber\\
   &&~~~~ -
    t_T \left( K^-,\, K^+,\, K^{*+},\, m_{a_0} \to M_{\rm inv}(\pi^0 \eta) \right) \cdot t_{K^+ K^-,\, \pi^0 \eta}(M_{\rm inv}(\pi^0 \eta)) \Big\}.~~~~
\end{eqnarray}

The production of the $f_0(980)$, which is related to the $\pi^+ \pi^-$ channel, proceeds in the same way. If we look into the $\pi^+ \pi^-$ decay channel, all we must do is to replace $\pi^0 \eta$ in Eqs.~\eqref{eq:ddGamma-1} and \eqref{eq:replace} by $\pi^+ \pi^-$, substituting $M_{\rm inv}(\pi^0 a_0) \to M_{\rm inv}(\pi^0 f_0)$ and
\begin{equation}
  \tilde{q}_\eta \to \tilde{q}_{\pi^-} =\frac{\lambda^{1/2} \left( M^2_{\rm inv}(\pi^+ \pi^-),\, m^2_{\pi^+}, \, m^2_{\pi^-} \right)}{2M_{\rm inv}(\pi^+ \pi^-)}.
\end{equation}

%%%%%%%%%%%%%%%%%%%%%%%%%%%%%%%%%%
\section{Results}
\label{sec:res}
\begin{figure}[bht]
\begin{center}
\includegraphics[scale=0.33]{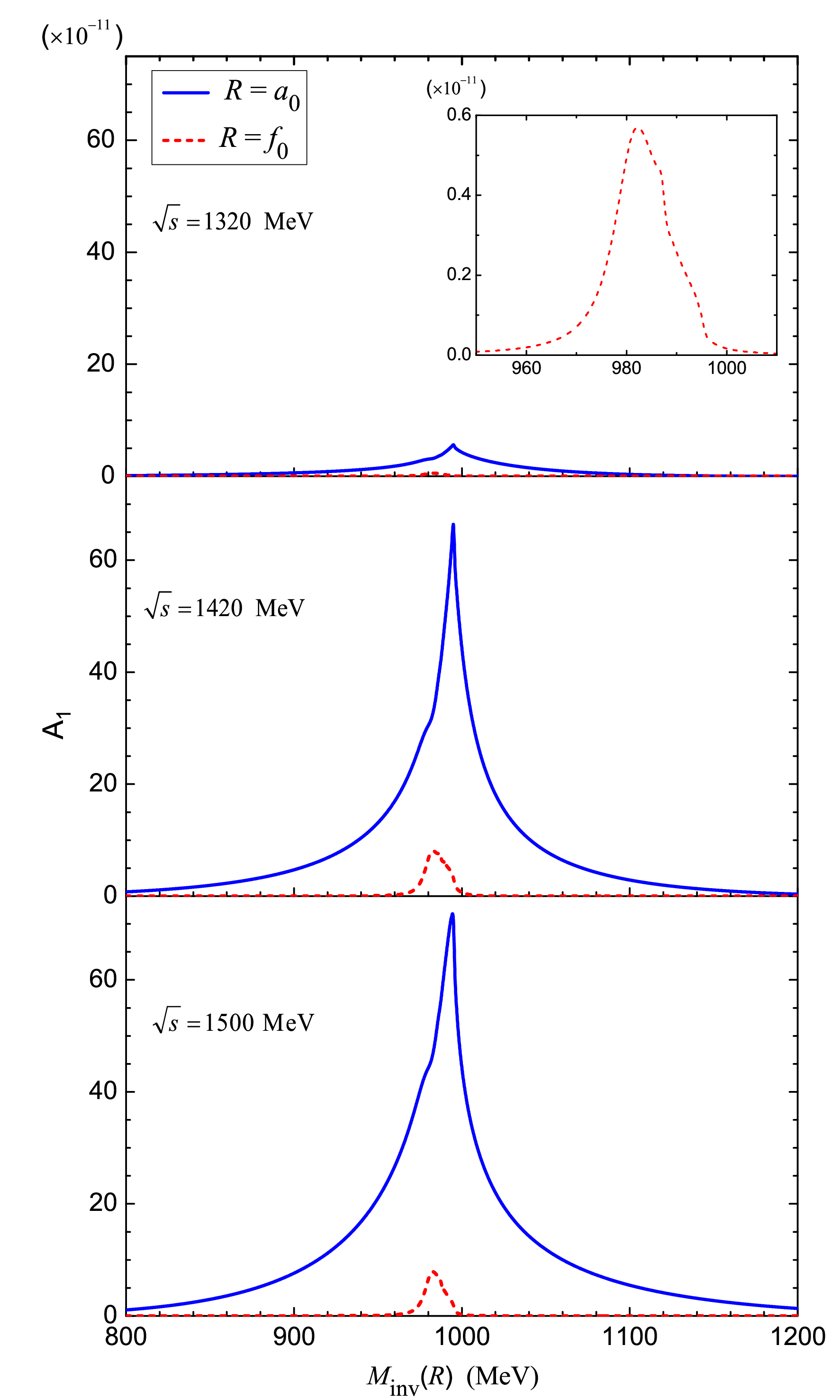}
\end{center}
\vspace{-0.3cm}
\caption{$\frac{1}{\Gamma_{\bar B_s^0}} \frac{{\rm d}^2 \Gamma_{\bar B_s^0 \to J/\psi \pi^0 \pi^0 \eta}}{{\rm d} M_{\rm inv} (\pi^0 a_0) \;{\rm d} M_{\rm inv} (\pi^0 \eta)}$ and
$\frac{1}{\Gamma_{\bar B_s^0}} \frac{{\rm d}^2 \Gamma_{\bar B_s^0 \to J/\psi \pi^0 \pi^+ \pi^-}}{{\rm d} M_{\rm inv} (\pi^0 f_0) \;{\rm d} M_{\rm inv} (\pi^+ \pi^-)}$ as functions of $M_{\rm inv}(\pi^0 \eta)$ or $M_{\rm inv}(\pi^+ \pi^-)$ for fixed value of $\sqrt{s}\equiv M_{\rm inv}(\pi^0 a_0)$ or $M_{\rm inv}(\pi^0 f_0)$ as 1320, 1420 and 1500 MeV, respectively.
Note: in this figure, the label of the longitudinal axis is ${\rm A}_1 = \frac{1}{\Gamma_{\bar B_s^0}} \frac{{\rm d}^2 \Gamma_{\bar B_s^0 \to J/\psi \pi^0 R}}{{\rm d} M_{\rm inv} (\pi^0 R) \;{\rm d} M_{\rm inv} (R)} [{\rm MeV}^{-2}]$. The inset magnifies the $M_{\rm inv}(\pi^+ \pi^-)$ distribution at fixed $\sqrt{s}=1320\, {\rm MeV}$.}
\label{Fig:3}
\end{figure}

As we have mentioned, we expect the TS to appear at $M_{\rm inv}(\pi^0 a_0)$ or $M_{\rm inv}(\pi^0 f_0) \approx 1424~{\rm MeV}$. In Fig.~\ref{Fig:3} we show the double
mass distribution
as a function of $M_{\rm inv} (R)$ (i.e. $M_{\rm inv}(\pi^0 \eta)$ or $M_{\rm inv}(\pi^+ \pi^-)$) for fixed values of $M_{\rm inv}(\pi^0 a_0)$ or $M_{\rm inv}(\pi^0 f_0)$. We take three values around the peak of the TS, 1320 MeV, 1420 MeV and 1500 MeV.

\begin{figure}[bpt]
\begin{center}
\includegraphics[scale=0.4]{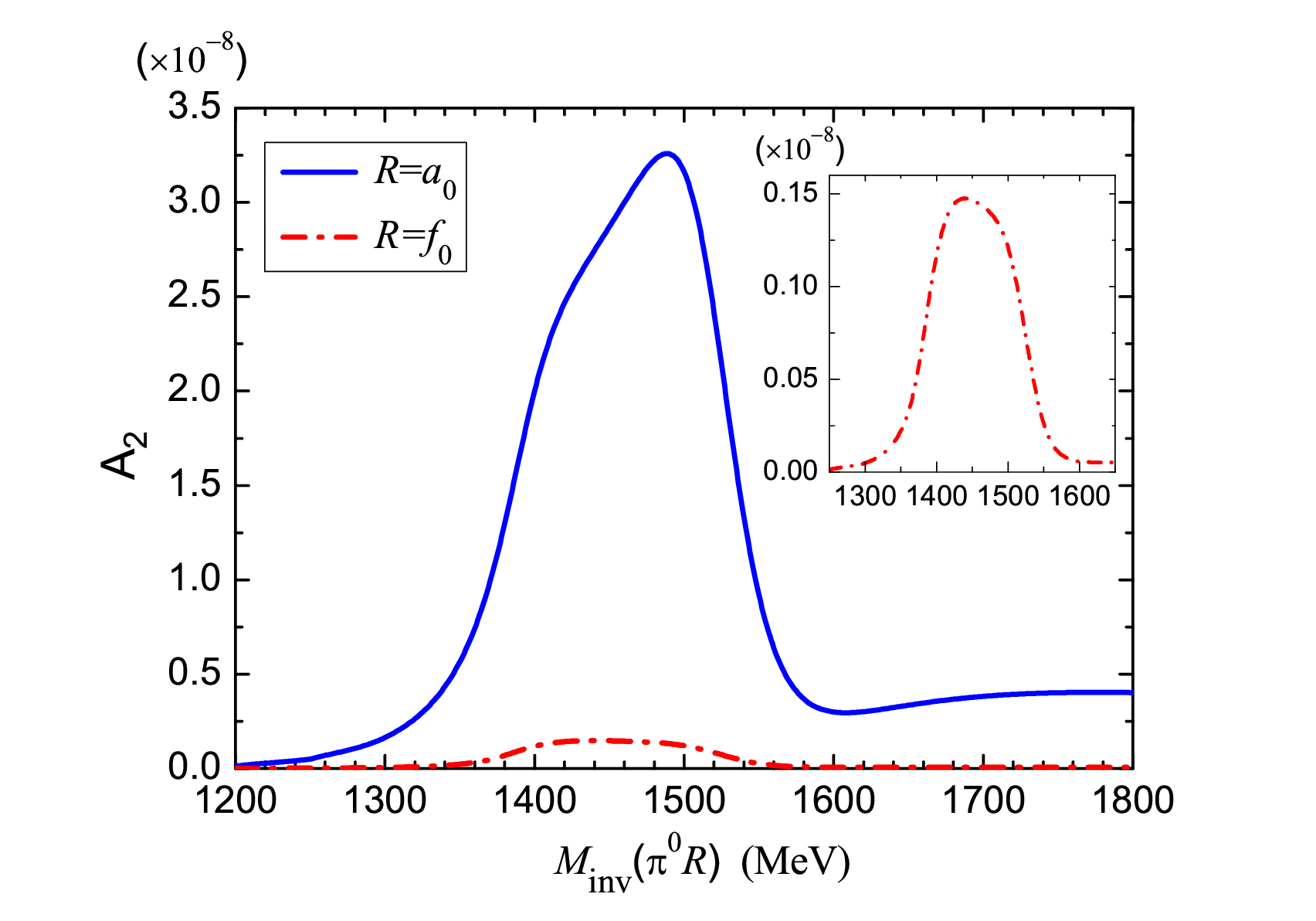}
\end{center}
\vspace{-0.7cm}
\caption{${\rm d} \Gamma / {\rm d} M_{\rm inv}(\pi^0 a_0)$ and ${\rm d} \Gamma / {\rm d} M_{\rm inv}(\pi^0 f_0)$ integrated over the respective $a_0$ and $f_0$ mass distributions (see text). Only the $\pi^+ \pi^-$ mode of $f_0$ and $\pi^0 \eta$ mode of $a_0$ are considered here. Note: in this figure, the label of the longitudinal axis is ${\rm A}_2 = \frac{1}{\Gamma_{\bar B_s^0}}\frac{{\rm d}\Gamma_{\bar B_s^0 \to J/\psi \pi^0 R}}{{\rm d}M_{\rm inv}(\pi^0 R)} [{\rm MeV}^{-1}]$. The inset magnifies the $\pi^0 f_0$ distribution.}
\label{Fig:4}
\end{figure}

As we can see for $M_{\rm inv}(\pi^0 R) \; (R=a_0, f_0)$ at 1420 MeV,
we get a large strength for $a_0$ production as well as $f_0$,
compared to the other two $M_{\rm inv}(\pi^0 R)$ masses,
which are away from the TS invariant mass.
The effect of the TS can be more clearly seen in Fig.~\ref{Fig:4},
where we have integrated the double mass distribution over
$M_{\rm inv} (R)$ (i.e. $M_{\rm inv}(\pi^0 \eta)$ or $M_{\rm inv}(\pi^+ \pi^-)$).
For the sake of comparison we have taken the range
$M_{\rm inv}(R) \in [\,950\, {\rm MeV},\, 1050\, {\rm MeV}\,]$.
The results of $\frac{{\rm d}\Gamma}{{\rm d}M_{\rm inv}(\pi^0 R)}$ are shown
in Fig.~\ref{Fig:4} and we observe that both the $\pi^0 a_0$ and
$\pi^0 f_0$ mass distributions have a clear peak around $M_{\rm inv}(\pi^0 R)=1420\, {\rm MeV}$.

In $\pi^0 a_0$ production there is a bump around $1420\, {\rm MeV}$, clearly attributable to the TS, while we also observe a neater peak around $1500\, {\rm MeV}$, whose origin is obviously the resonance shape of the original $K^* \bar K$ production shown in Fig.~\ref{fig:mass-dist-primary}. Curiously, in the $\pi^0 f_0$ production the situation is reversed and the peak appears at $1420\, {\rm MeV}$, while at $1500\, {\rm MeV}$ there is just a soft bump. This means that the TS is very effective at enhancing the isospin-forbidden $\pi^0 f_0$ mode.

\begin{figure}[t!]
\begin{center}
\includegraphics[scale=0.35]{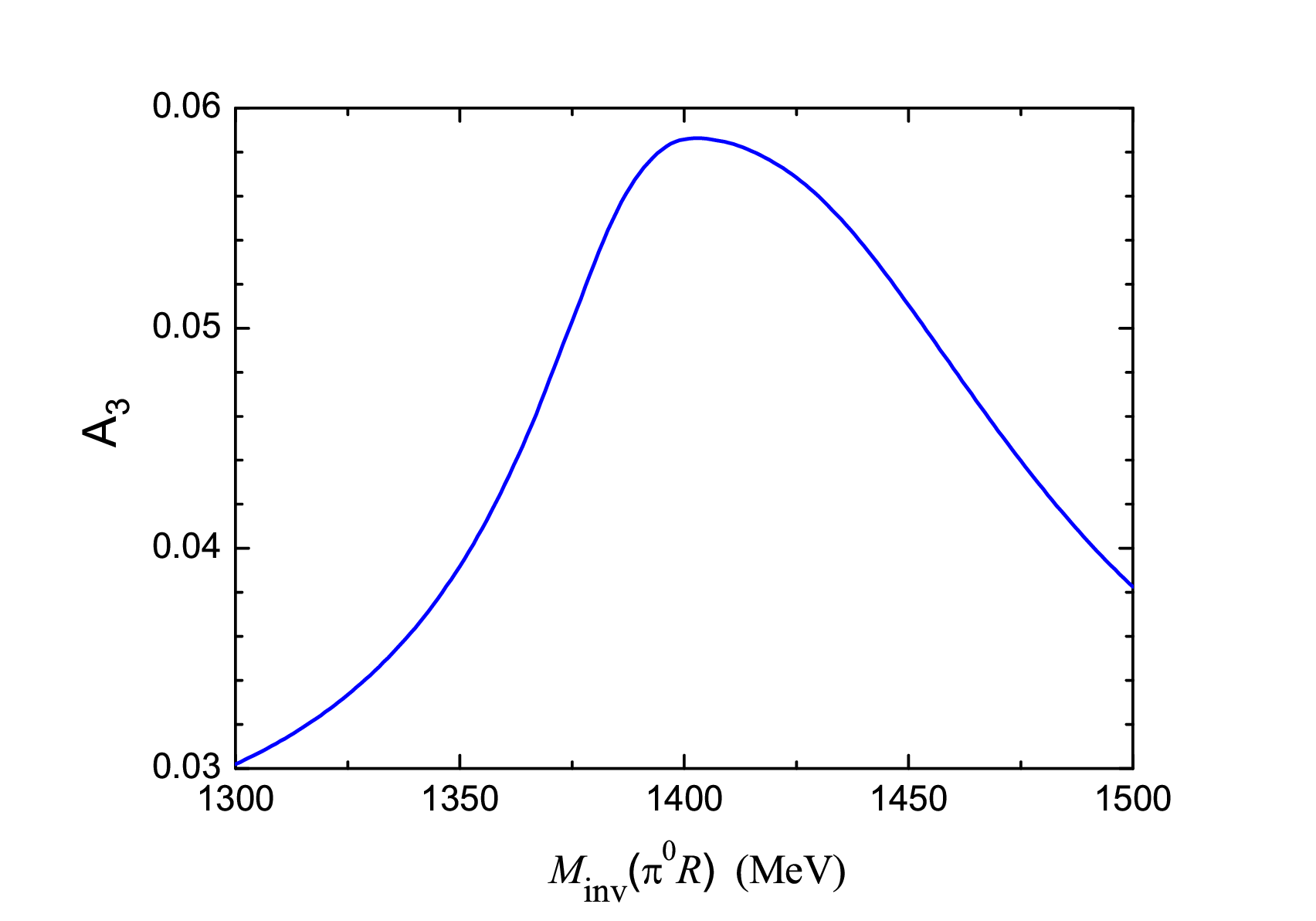}
\end{center}
\vspace{-0.7cm}
\caption{Ratio of ${\rm d} \Gamma / {\rm d} M_{\rm inv}(\pi^0 a_0)$ and ${\rm d} \Gamma / {\rm d} M_{\rm inv}(\pi^0 f_0)$ as a function of $M_{\rm inv} (\pi^0 R) \; (R=f_0, a_0)$. Note: in this figure, the label of the longitudinal axis is ${\rm A}_3 = \frac{{\rm d}\Gamma_{\bar B_s^0 \to J/\psi \pi^0 f_0}}{{\rm d}M_{\rm inv}(\pi^0 f_0)} {\Big/ } \frac{{\rm d}\Gamma_{\bar B_s^0 \to J/\psi \pi^0 a_0}}{{\rm d}M_{\rm inv}(\pi^0 a_0)} $.}
\label{Fig:5}
\end{figure}

From Fig.~\ref{Fig:4}, we can also take the ratio of $\frac{{\rm d}\Gamma}{{\rm d}M_{\rm inv}(\pi^0 f_0)}$ versus $\frac{{\rm d}\Gamma}{{\rm d}M_{\rm inv}(\pi^0 a_0)}$, which we show in Fig.~\ref{Fig:5}, and we see that this ratio also peaks around the mass of the TS, although shifted a bit to lower invariant masses.
The resonant shape of the $K^* \bar K$ production has no role in this ratio, because the factor $|F(M_{\rm inv}(\pi^0 R))|^2$ is the same in the two distributions and cancels in the ratio.
In other words, the TS enhances the isospin-violating mode $\pi^0 f_0$ in absolute terms, but also relative to the isospin-allowed $\pi^0 a_0$ mode.

\begin{figure}[tbhp]
\begin{center}
\includegraphics[scale=0.43]{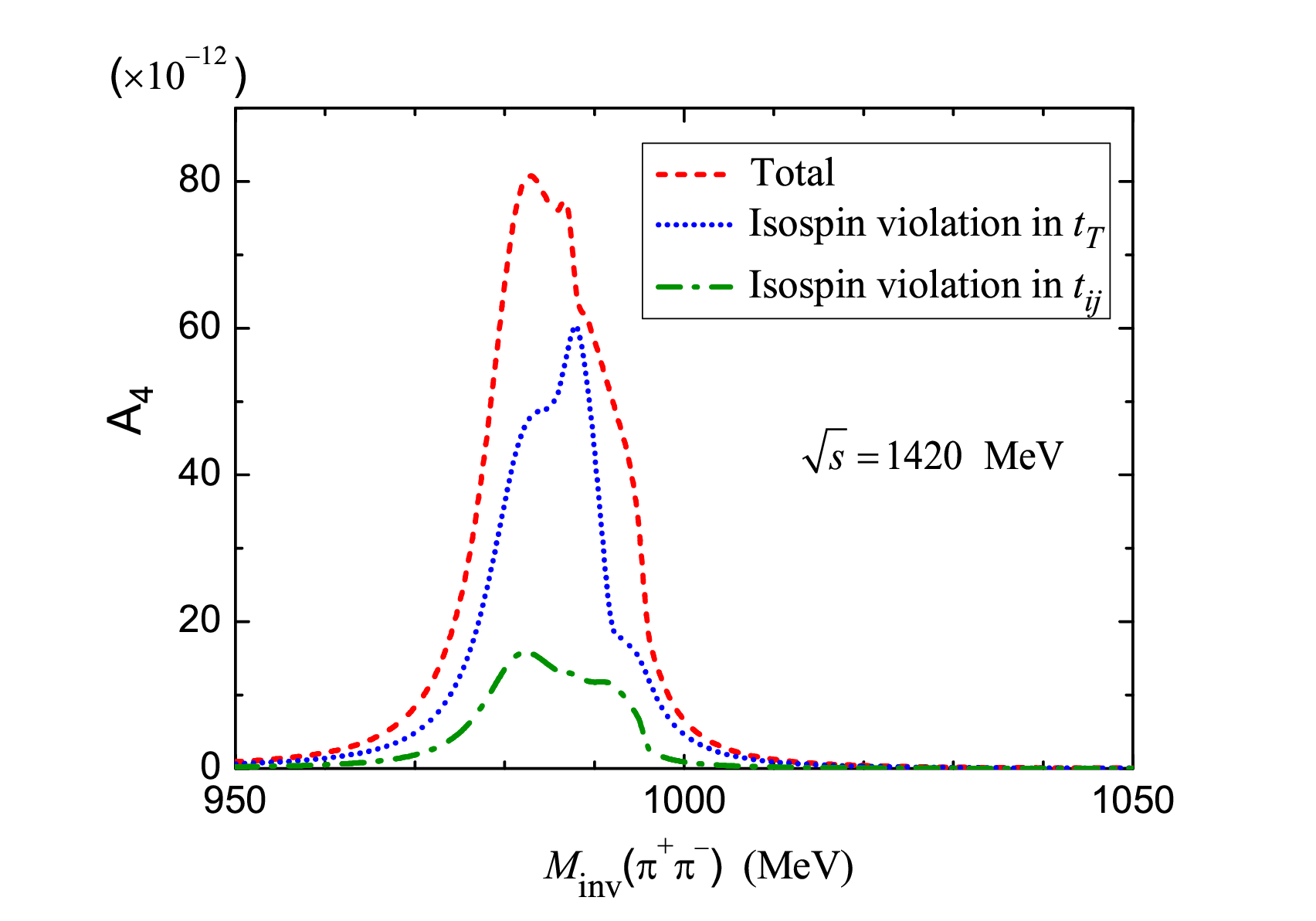}
\end{center}
\vspace{-0.7cm}
\caption{$\frac{1}{\Gamma_{\bar B_s^0}} \frac{{\rm d}^2 \Gamma_{\bar B_s^0 \to J/\psi \pi^0 \pi^0 \eta}}{{\rm d} M_{\rm inv} (\pi^0 f_0) \;{\rm d} M_{\rm inv} (\pi^+ \pi^-)}$ for fixed $M_{\rm inv} (\pi^0 f_0)=1420\, {\rm MeV}$, for two cases, isospin violation only in $t_T$ and isospin violation only in $K\bar K \to \pi^+ \pi^-$. Note: in this figure, the label of the longitudinal axis is ${\rm A}_4 = \frac{1}{\Gamma_{\bar B_s^0}} \frac{{\rm d}^2 \Gamma_{\bar B_s^0 \to J/\psi \pi^0 f_0}}{{\rm d} M_{\rm inv} (\pi^0 f_0) \;{\rm d} M_{\rm inv} (\pi^+ \pi^-)} [{\rm MeV}^{-2}]$.}
\label{Fig:6}
\end{figure}

It is interesting to see the sources of isospin violation. They are tied to the differences of the $K^0, K^+$ masses, but they influence both $t_T$ in the triangle singularity as well as the two-body scattering matrices $t_{ij}$ for $K \bar K \to \pi^0 \eta$ and $K\bar K \to \pi^+ \pi^-$. To show the effects independently, we take the middle mass $M_{\rm inv}(\pi^0 R)$ in Fig.~\ref{Fig:3} and show the $\pi^0 f_0$ production in two cases: One assuming equal $K$ masses in the evaluation of the $K\bar K \to \pi^0 \eta, \pi^+ \pi^-$ amplitudes (isospin symmetry in the meson scattering amplitudes) and keeping different $K$ masses in the triangle loop evaluation, $t_T$, and another case in which we take equal $K$ masses in $t_T$ but different masses in the meson amplitudes. The results can be seen in Fig.~\ref{Fig:6}, where the ``Total'' line contains isospin violation both in $t_T$ and $t_{ij}$, same as in Fig.~\ref{Fig:3}.
We can see that both effects are important and they add to the total amplitude producing $\pi^0 f_0$. These results are similar to those found in the study of the $\chi_{c1} \to \pi^0 f_0(980) (\pi^+ \pi^-)$ and $\chi_{c1} \to \pi^0 a_0(980) (\pi^0 \eta)$ in Ref.~\cite{melavini}.
In the figure one can observe two structures to the right of the invariant mass distribution corresponding to the $K^+ K^-$ and $K^0 \bar K^0$ thresholds.

\begin{figure}[t]
 \centering
 \includegraphics[width=5cm]{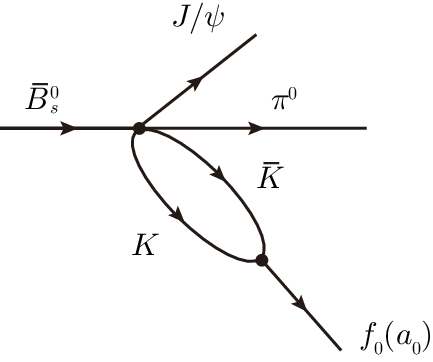}
 \caption{Effective mechanism resulting from taking the $K^*$ far off
 shell in the diagram of Fig.~\ref{Fig:2},
 reducing the $K^*$ mass to 300~MeV.}
 \label{fig_new}
\end{figure}

\begin{figure}[t]
 \centering
 \includegraphics[scale=0.4]{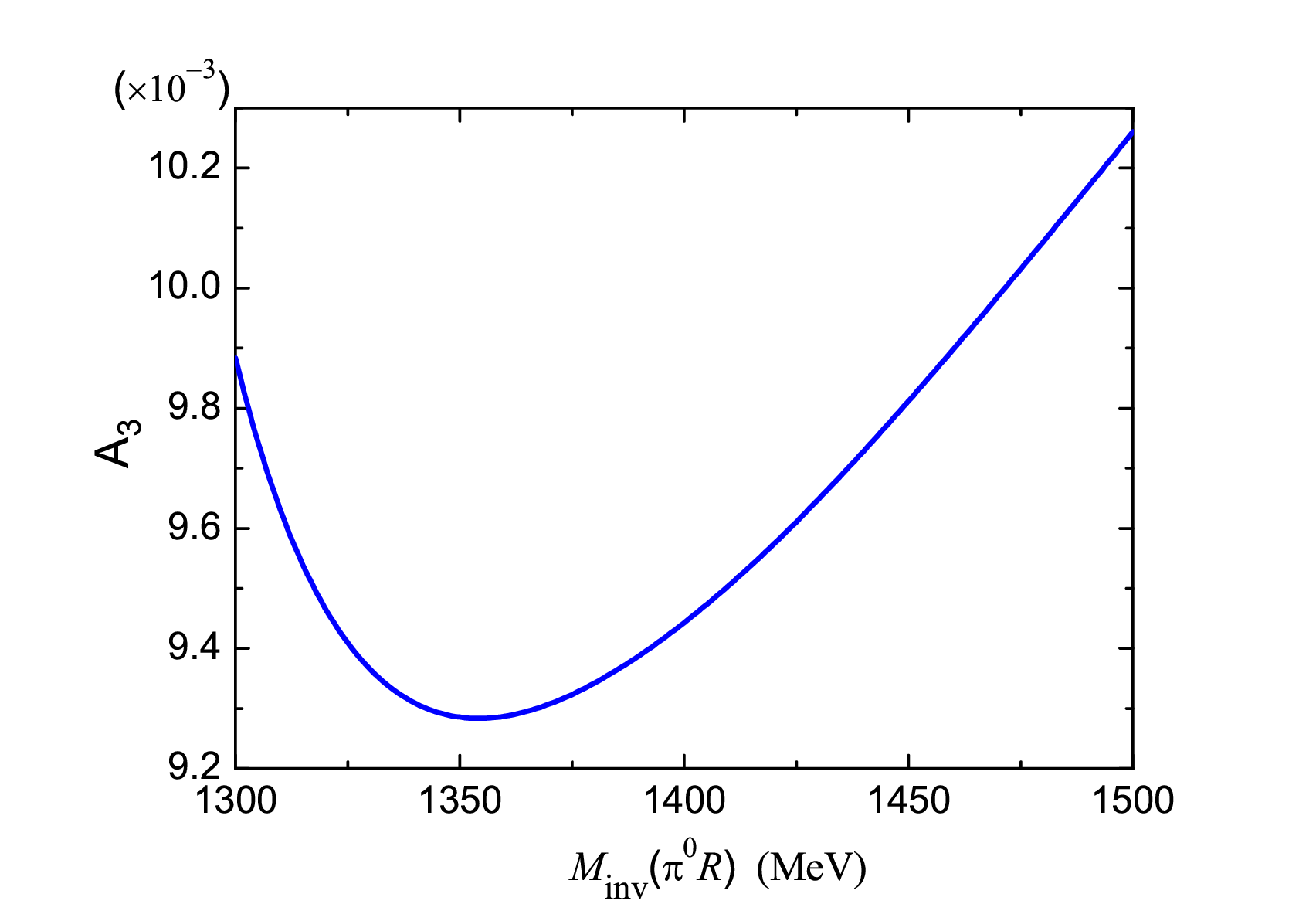}
 \caption{Ratio of ${\rm d} \Gamma / {\rm d} M_{\rm inv}(\pi^0 a_0)$ and ${\rm d} \Gamma / {\rm d} M_{\rm inv}(\pi^0 f_0)$ as a function of $M_{\rm inv} (\pi^0 R) \; (R=f_0, a_0)$, taking $m_{K^*}=300$~MeV, which makes the TS disappear. Note: in this figure, the label of the longitudinal axis is ${\rm A}_3 = \frac{{\rm d}\Gamma_{\bar B_s^0 \to J/\psi \pi^0 f_0}}{{\rm d}M_{\rm inv}(\pi^0 f_0)} {\Big/ } \frac{{\rm d}\Gamma_{\bar B_s^0 \to J/\psi \pi^0 a_0}}{{\rm d}M_{\rm inv}(\pi^0 a_0)} $.}
 \label{fig_newnew}
\end{figure}

It is interesting to compare the behavior of Fig.~\ref{Fig:5} with
what we should expect if there is no triangle singularity.
For this purpose we use the same formalism but artificially change the
mass of the $K^*$ to 300~MeV and its width to zero.
This guarantees that when $K$ and $\bar{K}$ are close to on-shell to
produce the $f_0$ or $a_0$, the $K^*$ is far off-shell and acts as a
point-like interaction.
Then we would have a mechanism as depicted in Fig.~\ref{fig_new}.
The result for the new ratio can be seen in Fig.~\ref{fig_newnew}.

The results are interesting. We can see that the ratio is practically constant between $1300\, {\rm MeV}$ and $1500\, {\rm MeV}$. It ranges from $9.3 \times 10^{-3}$ to $10.2 \times 10^{-3}$ in that range, while in Fig.~\ref{Fig:5} it changes in a factor two in that range. Note also that in Fig.~\ref{Fig:5} the results are about a factor six bigger than in Fig.~\ref{fig_newnew}, indicating the importance of the TS inducing the isospin-violating mode of $\pi^0 f_0$.

Finally, in order to estimate the total rate for $\bar B_s^0 \to J/\psi \pi^0 f_0$ and $\bar B_s^0 \to J/\psi \pi^0 a_0$, we integrate $\frac{{\rm d}\Gamma}{{\rm d}M_{\rm inv}(\pi^0 R)}$ in Fig.~\ref{Fig:4} over the $\pi^0 R$ invariant mass in the range $[\,1200 \,{\rm MeV}, \,1600\,{\rm MeV}\,]$ of invariant masses of the figure and we find
\begin{equation}\label{eq:ratef0}
  {\rm Br}(\bar B_s^0 \to J/\psi \pi^0 f_0,\, f_0 \to \pi^+ \pi^-) = 2.2 \times 10^{-7},
\end{equation}
\begin{equation}\label{eq:ratea0}
  {\rm Br}(\bar B_s^0 \to J/\psi \pi^0 a_0) = 4.9 \times 10^{-6}.
\end{equation}
If we take into account the $\pi^0 \pi^0$ decay channel of the $f_0(980)$, which is one half of the $\pi^+ \pi^-$,
\begin{equation}\label{eq:ratef02}
  {\rm Br}(\bar B_s^0 \to J/\psi \pi^0 f_0) = 3.3 \times 10^{-7}.
\end{equation}
These rates are within present observation capability at LHCb.

\section{Conclusions}
\label{sec:conc}

We have made a study of the $\bar B_s^0 \to J/\psi \pi^0 f_0(980)\, (a_0(980))$ decay which proceeds via a triangle mechanism in which there is first the decay $\bar B^0_s \to J/\psi K^* \bar K$ or $\bar B^0_s \to J/\psi \bar K^* K$ and posterior fusion of $K\bar K$ to give the $f_0(980)$ or $a_0(980)$ resonance. The primary process at quark level is $\bar B_s^0 \to J/\psi\, s\bar s$, with the $s\bar s$ hadronizing into $K^* \bar K - \bar K^* K$, which guarantees isospin $I=0$ for this combination. This means that the isospin-allowed $\pi^0 R$ ($R=f_0,\, a_0$) final state is $\pi^0 a_0$, while the $\pi^0 f_0$ mode is isospin-suppressed. Yet, the explicit consideration of the $K^+, K^0$ different masses gives a contribution for $J/\psi \pi^0 f_0(980)$ at the end, with a shape for the $f_0(980)$ in the $\pi^+ \pi^-$ mass distribution tied to the difference of masses of $K^+, K^0$ and, hence, much narrower than the standard $f_0(980)$ shape seen in the isospin-allowed modes. This shape and strength are tied to the dynamically generated nature of the $f_0(980)$ and $a_0(980)$ as coming from the interactions of pseudoscalar mesons.

The shape obtained for this isospin-suppressed mode is in agreement with other experiments where the $f_0$ is also obtained with isospin-violating mechanisms. The novelty in the reaction proposed is that the triangle mechanism develops a triangle singularity at an invariant mass $M_{\rm inv} (\pi^0 f_0)$ of about 1420 MeV. Around this invariant mass the production of both the $J/\psi \pi^0 f_0$ and $J/\psi \pi^0 a_0$ modes are enhanced, and more notably the ratio of the $J/\psi \pi^0 f_0$ to $J/\psi \pi^0 a_0$ production also shows a peak around the triangle singularity point. This evidences the role of this triangle singularity in reinforcing isospin violation in the reaction. We also showed that the isospin-violating amplitude has two sources, one from the consideration of the different $K$ masses in the triangle loop, and the other one from the isospin violation in the meson-meson amplitudes, coming again from the consideration of different meson masses in the coupled channels unitary approach used to generate these amplitudes.

Using experimental input from the $B_s^0 \to J/\psi K^* \bar K + c.c.$ decay, we can make absolute predictions for the branching fractions of $\bar B_s^0 \to J/\psi \pi^0 f_0(980)\, (a_0(980))$ and find them within measurable range.

The predictions made, and their accessibility within present experimental facilities, should give a strong motivation to perform these experiments, which will provide valuable information on the nature of the low lying scalar mesons.

\vspace{-0.3cm}
\begin{acknowledgments}
One of us, S. Sakai, wishes to thank the Generalitat Valenciana in the
program Prometeo II-2014/068 for support.
This work is partly supported by the National Natural Science Foundation of China (Grants No. 11565007, 11747307, 11647309, 11735003 and 11475227) and the Youth Innovation Promotion Association CAS (No. 2016367).
This work is also partly supported by the Spanish Ministerio de Economia y Competitividad
and European FEDER funds under the contract number FIS2011-28853-C02-01, FIS2011-28853-C02-02, FIS2014-57026-REDT, FIS2014-51948-C2-1-P, and FIS2014-51948-C2-2-P, and the Generalitat Valenciana in the program Prometeo II-2014/068.
\end{acknowledgments}

%\newpage

  \end{document}